\documentclass[twocolumn,english,groupedaddress, superscriptaddress]{revtex4-1}
\usepackage[utf8]{inputenc}
\usepackage{amssymb}
\usepackage{amsmath}
\usepackage{mathrsfs}
\usepackage{graphicx}
\usepackage{color}

\begin{document}

\title{Curing Braess' Paradox by Secondary Control in Power Grids}

\author{Eder Batista Tchawou Tchuisseu}

\affiliation{Instituto de Física Interdisciplinar y Sistemas Complejos, IFISC (CSIC-UIB), Campus Universitat Illes Balears,
E-07122 Palma de Mallorca, Spain}

\author{Damià Gomila}

\affiliation{Instituto de Física Interdisciplinar y Sistemas Complejos, IFISC (CSIC-UIB), Campus Universitat Illes Balears,
E-07122 Palma de Mallorca, Spain}

\author{Pere Colet}

\affiliation{Instituto de Física Interdisciplinar y Sistemas Complejos, IFISC (CSIC-UIB), Campus Universitat Illes Balears,
E-07122 Palma de Mallorca, Spain}

\author{Dirk Witthaut}

\affiliation{Forschungszentrum J\"ulich, Institute for Energy and Climate Research
- Systems Analysis and Technology Evaluation (IEK-STE), 52428 J\"ulich,
Germany}

\affiliation{Institute for Theoretical Physics, University of Cologne, 50937 K\"oln,
Germany}

\author{Marc Timme}

\affiliation{Chair for Network Dynamics, Center for Advancing Electronics Dresden (cfaed) and Institute for Theoretical Physics,
Technical University of Dresden, 01062 Dresden, Germany}

\affiliation{Network Dynamics, Max Planck Institute for Dynamics and Self-Organization
(MPIDS), 37077 Göttingen, Germany}

\author{Benjamin Sch\"afer}

\affiliation{Chair for Network Dynamics, Center for Advancing Electronics Dresden (cfaed) and Institute for Theoretical Physics,
Technical University of Dresden, 01062 Dresden, Germany}

\affiliation{Network Dynamics, Max Planck Institute for Dynamics and Self-Organization
(MPIDS), 37077 Göttingen, Germany}

\begin{abstract}
Robust operation of power transmission grids is essential for most of today's technical infrastructure and our daily life.
Adding renewable generation to power grids requires grid extensions and sophisticated control
actions on different time scales to cope with short-term fluctuations and long-term power imbalance.
Braess' paradox constitutes a counterintuitive collective phenomenon that occurs if adding new transmission line capacity to a network increases loads on other lines, effectively reducing the system's performance and potentially even entirely removing its operating state. Combining simple analytical considerations with numerical investigations on a small sample network, we here study dynamical consequences of secondary control in AC power grid models.  We demonstrate that sufficiently strong control not only implies dynamical stability of the system but may also cure Braess' paradox. Our results highlight the importance of demand control in conjunction with grid topology for stable operation and reveal a new functional benefit of secondary control.
\end{abstract}

\maketitle

\section{Introduction}
Modern electrical power grids are complex  interconnected networks in which supply and demand  have to match at all times since the grid itself cannot store any energy \cite{Kundur1994,Brummitt2013}. To guarantee this match, different economic mechanisms, like day-ahead and intra-day markets are used \cite{Kovacevic2013}. 
For unscheduled mismatches, e.g. random fluctuations \cite{Schaefer2017a}, disturbances or extreme weather, faster control mechanisms are required \cite{Machowski2011}. Such control actions become increasingly important due to the rising share of renewable generation integrated into the grid \cite{Boyle2004,Sims2011,Ueckerdt2015}.
Control mechanisms are ordered by their time scale on which they act: Suppose a power plant has to unexpectedly shut down and all of a  sudden there is a shortage of 
energy in the system. The first second of the disturbance is mainly uncontrolled, i.e. energy is drawn from the spinning reserve of the generators.
Within the next seconds, the primary control sets in to stabilize the frequency and prevents a large drop. 
To restore the frequency back to its nominal value of 50 Hz or 60 Hertz, secondary control is necessary \cite{Machowski2011}.
However, in many recent studies on power system dynamics and stability, the effects of control are completely neglected or only primary control is considered \cite{Filatrella2008,Rohden2012,Motter2013a,Doerfler2014,Manik2014,
Schaefer2015,Schaefer2016}. Including secondary control might be crucial when determining stability conditions.
Even in cases where secondary control is modeled explicitly \cite{Weitenberg2017}, its stability properties and interaction with the network topology are not fully investigated.\\
Nonetheless, grid topology and control mechanisms have to adapt within the next years to cope with the spatially distributed and fluctuating renewable generation, which may be far away from consumers \cite{Ipakchi2009}. Grid adaptation includes additional transmission lines and increasing capacity of existing lines \cite{50Hertz2012,Fuersch2013}. Contrary to expectations, not all added lines are beneficial to the stability of a grid. Instead, adding some lines may cause the grid to lose its operating state via  Braess' paradox, which was initially discovered for transportation networks \cite{Braess1968} but may also occur in power grids  \cite{Witthaut2012,Witthaut2013,Nagurney2016,Coletta2016,Fazlyaba2015}.
\\
Here, we present a dynamical analysis on the effectiveness and limitations of an implementation of secondary control that depends on the absolute voltage phase angle $\theta$ of a synchronous machine.
We dynamically show how a simple implementation of secondary control restores a power grid with a power mismatch back to the nominal frequency. 
Furthermore, we investigate the stability of a grid with secondary control as a function of the network topology. In particular, we study a setting in which secondary control prevents Braess' paradox. We find that Braess' paradox is reliably avoided if all nodes are controlled. However, controlling only generators still allows Braess' paradox as before, thereby highlighting the importance of demand-side control in future grids.
\\
This article is structured as follows. 
First, we present a simple model of the dynamics of the electric power network in the presence of secondary control in Section II.
The generators and consumers  are modeled as synchronous machines and the control is modeled by using Load Frequency Control (LFC) and  Automatic
 Generation Control (AGC) \cite{Mallorca2015}.
 Next, we investigate the dynamics of the grid including secondary control and demonstrate its capability to stabilize a grid with a power mismatch in Section III.
 Section IV investigates Braess' paradox when applying secondary control. 
We close the paper with a discussion on the impact of our results on current and future power grids.

\section{Mathematical modeling  of the electric power system}
Most of today's electric power is generated by three phase synchronous generators, e.g. driven by steam or hydro turbines, which output an alternating current (AC) \cite{Anderson2008}. 
The electric power grid may then be modeled as an interconnected network consisting of nodes linked by power transmission lines (links). Each node can be interpreted as a local area including power generation and consumption with net mechanical power input $P^m_i$ being negative for effective consumer regions, e.g. urban areas, and positive for effective generators. We model each node by 
the well-known \textit{swing equation} \cite{Kundur1994,Filatrella2008,Machowski2011,Rohden2012,Schaefer2015}, which gives the equation of motion in the rotating frame of each node $\textit{i}$ as
 \begin{equation}  
  \begin{array}{ll}
      \dot{\mathbf{\theta}}_i=\omega_i, \\
      \dot{\omega}_{i} =\frac{\omega_R}{2HP^\text{Gmax}_i}(P^\text{m}_{i}(\omega_{i})-P^\text{e}_i(\theta_{i},\omega_{i})).
   \end{array}
\label{eq1} \end{equation}
The state of node \textit{i} is characterized in the co-moving reference frame  by the voltage phase angle $\theta_i$ and the angular velocity deviation $\omega_i$.
$P^\text{e}_i$ represents the total power consumed and transmitted at node $i$. Finally, $\omega_R$=$2\pi \times 50$~Hz or $60$~Hz is the reference angular velocity of the grid and $H$ is the inertia constant of the generator with a power rate $P^\text{Gmax}_i$. Due to the choice of reference frame, $\omega_i=0$ implies a frequency of 50 or 60 Hz.

The power grid is subject to fluctuations, e.g.  due to changing demand, volatile generation of renewables or trading \cite{Heide2010,Milan2013,Schaefer2017a}. To cope with these fluctuations, the grid is controlled on multiple time scales with primary control being the fastest, followed by secondary control. The primary frequency control is typically done by a few dedicated power plants, which adjust
their mechanical power output proportional to the angular velocity deviation $\omega_i$ \cite{Machowski2011,Anderson2008}. 
Secondary control is then applied through Automatic Generation Control (AGC) to restore the frequency, using spinning and non-spinning reserves.
We model this frequency regulation as a Proportional Derivative (PD) control \cite{Mallorca2015}. Alternatives include Proportional (P) control, Proportional Integral (PI) control or Proportional Integral Derivative (PID) control \cite{wang2016control,Simpson-Porco2012,hammid2016load}.

Given this implementation of primary and secondary control and assuming that the voltage amplitude is constant, we model each machine $i$ by means of the swing equation \cite{Saadat2002, Mallorca2015}, which includes a Kuramoto-like coupling between the machines \cite{Manik2014}:
 \begin{equation}  
    \begin{array}{ll}
      \dot{\mathbf{\theta}}_i={\omega}_{i},\\
      \dot{\omega}_i=-\alpha_{i}{\omega}_{i}+{P_{i}}-\sum\limits_{j=1}^n K_{ij}\sin(\theta_i - \theta_j) + \Delta P_{c_i}(t),\\
     \tau_{i} \dot{\Delta P_{c_i}}=-\Delta P_{c_i}(t)-[\beta_{i}{\omega}_{i} (t)+\gamma_{i}\theta_{i}(t)],
     \end{array}
 \label{eq7} 
\end{equation}
where  $\alpha_i$ is a damping constant due to losses and damper windings and $K_{ij}$ is proportional to the susceptance of line $(i,j)$ and gives the capacity of a line. ${P_{i}}$ is the effective power fed into the grid or consumed at node $i$. ${P_{i}}$ is positive for generators, while it is negative for consumers.
Finally, $\Delta P_{c_i}$ is the control power with time constant $\tau_{i}$, derivative and proportional control gain $\beta_i$ and $\gamma_i$  respectively. The term $\gamma_i$ gives essentially the magnitude of the secondary control while $\beta_i$  determines the magnitude of the primary control.
Eq. \eqref{eq7} has the form of the well-known 2nd-order Kuramoto model, which has been used for example in \cite{Filatrella2008,Grzybowski2016} without control to describe the dynamics of the power grid.

In the remainder of this article, we set the parameter $\tau_{i}$=0, meaning that the control acts instantaneously. This approximation does not affect the final steady state of the system, which we are mainly interested in, simplifying the model considerably.
The time constant $\tau_{i}$ only changes the frequency of the oscillations during the transient dynamics. Thereby, we can solve the equation for $\Delta P_c$ and insert it into the equation of $\dot\omega_i$. In addition, since the damping $\alpha_i$ and primary control $\beta_i$ play a similar dynamical role, we absorb any contribution from $\beta_i$ into $\alpha_i$, effectively setting $\beta_i=0$.  With that, our equation of motion for each machine reads
\begin{equation}  
    \begin{array}{ll}
      \dot{\mathbf{\theta}}_i={\omega}_{i},\\
      \dot{\omega}_i=-\alpha_{i}{\omega}_{i}  - \gamma_{i}\theta_{i} + {P_{i}}-\sum\limits_{j=1}^n K_{ij}\sin(\theta_i - \theta_j).\\
     \end{array}
 \label{eq8} 
\end{equation}

Throughout this article, we will initialize numerical simulations of eq. \eqref{eq8} using $\theta_{i}(0)=0$ and $\omega_{i}(0)=0$ for all nodes as initial conditions.

\section{Steady state analysis and stability condition}
The power grid is in a steady state when all rotatory machines are phase-locked, i.e, have the same frequency, which ideally is the reference frequency of $f_\text{R}=50$~Hz or 60~Hz \cite{Doerfler2013}. 
To derive the stability conditions of the synchronous state with respect to small perturbations, we linearize Eq. \eqref{eq8} around  a steady state $(\theta_{i}^{*},\omega_{i}^{*})$.
We denote small perturbations around the fixed point as $\theta_{i}$=$\theta_{i}^{*}$+$\delta\theta_{i}$ and $\omega_{i}$=$\omega_{i}^{*}$+$\delta\omega_{i}$
and define $\mathbf{X_1}$ and $\mathbf{X_2}$, as the $n$-dimensional vectors of $\delta\theta_{i}$
and $\delta{\omega}_{i}$, respectively. 
Linearizing \eqref{eq8} yields

\begin{equation}  
 \begin{array}{ll}
      \dot{\mathbf{X}}_1=\mathbf{X_2},\\
      \dot{\mathbf{X}}_2=-(\mathbf{L}+\mathbf{\Gamma})\mathbf{X_1}-\mathbf{A}\mathbf{X_2},
     \end{array}
 \label{eq7a} 
\end{equation}
where $\mathbf{\Gamma}$ and $\mathbf{A}$ are diagonal matrices with elements $\Gamma_{ii}$= $\gamma_i$ and $A_{ii}$= $\alpha_i$ respectively, representing
the control and the damping matrix. Matrix $\mathbf{L}$=$(L_{ij})$ is a Laplacian matrix of the network topology, defined as
\begin{equation}
L_{ij}=\left\{ \begin{array}{ll} 
		-K_{ij}\cos(\theta^*_{i}-\theta^*_{j}), 
		& i \ne j,\\
        -\sum\limits_{l \ne i}^n L_{il},
        & i=j.
          \end{array}
\right.
\end{equation}
We diagonalize the Laplacian matrix $\mathbf{L}$ by substituting $\mathbf{Y_{1}}=\mathbf{M}^{-1}\mathbf{X_{1}}$, $\mathbf{Y_{2}}=\mathbf{M}^{-1}\mathbf{X_{2}}$, where $\mathbf{M}$ is the matrix composed of the eigenvectors of $\mathbf{L}$ such that
that $\mathbf{J}=\mathbf{MLM}^{-1}$ is the diagonalized matrix composed by the eigenvalues $\mu_{j}$.
Eq. \eqref{eq7a} can be rewritten as
\begin{equation}
\frac{\text{d}}{\text{d}t}\left[ {\begin{array}{cc}
 \mathbf{Y_{1j}}\\
 \mathbf{Y_{2j}}
\end{array} } \right]
= \left[ {\begin{array}{cc}
 0 & 1 \\
 -\mu_{j}-\gamma_j & -\alpha_j
\end{array} } \right]\left[ {\begin{array}{cc}
 \mathbf{Y_{1j}}\\
 \mathbf{Y_{2j}}
\end{array} } \right].
\end{equation}
\\
Without secondary control, i.e., $\gamma_i=0$, this dynamical system exhibits a single zero eigenvalue that does not determine the overall system stability but arises because the stability is only defined up to an arbitrary phase shift, i.e., we could replace all phases by different ones by 
adding a constant everywhere \cite{Manik2014}
\begin{equation} 
\theta_i \rightarrow \tilde\theta_i=\theta_i+\text{const}.
\end{equation}
However, for the general case of $\gamma_i>0$, the synchronous state of the system is stable if and only if the real parts of all  lyapunov exponents $\lambda_j$
\begin{equation}
 \lambda_{{j}{\pm}}=-\frac{\alpha_j}{2} \pm \frac{1}{2}\sqrt{{\alpha_j}^2-4(\mu_{j}+\gamma_j)}
 \label{eq:eigenvalue as function of laplacian}
\end{equation} are negative.
We assume symmetric coupling $K_{ij}=K_{ji}$; thereby guaranteeing real eigenvalues ${\mu_j}$.

\begin{figure*}[t]
 \centering
    \includegraphics[scale=0.45] {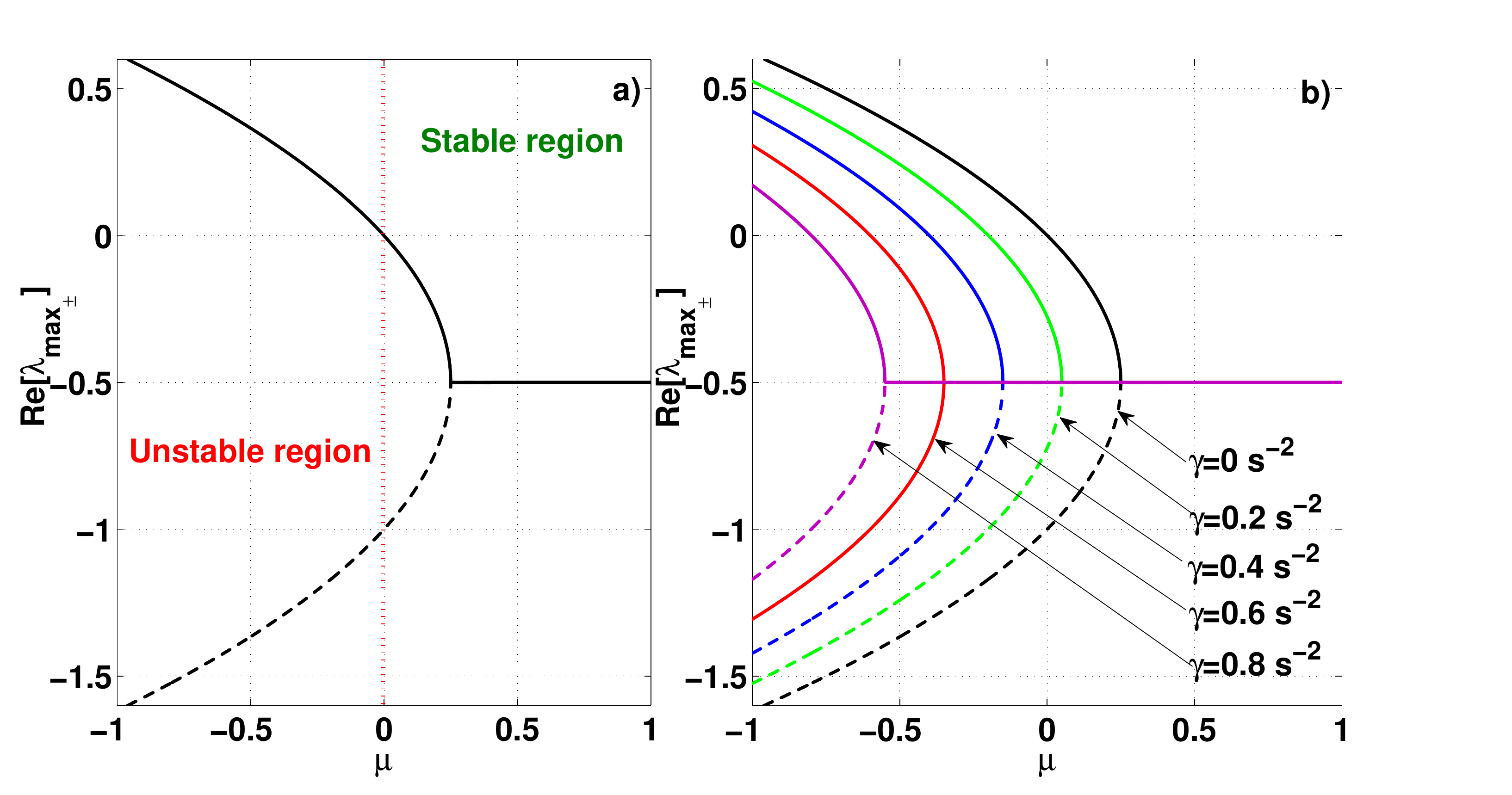} 
  \caption{  
\textbf{Secondary control extends the stable operation as a function of the topology ($\mu$).}  
 We plot the real part of the largest eigenvalue $\lambda_{+}$ (continuous lines) and $\lambda_{-}$ (dashed lines) as functions of the Laplacian eigenvalue $\mu$, see Eq. \eqref{eq:eigenvalue as function of laplacian}. 
 We assume that the control is homogeneous throughout the network, i.e., $\gamma_j=\gamma$.
  (a) Without control, $\gamma=0$, the system becomes unstable as soon as the Laplacian eigenvalue $\mu$ becomes negative, as then $Re(\lambda)>0$. Note that Eq. \eqref{eq:eigenvalue as function of laplacian} starts having two solutions as soon as $\alpha^2=4(\mu+\gamma)$.
  (b) With increasing control, $\gamma>0$, the region of stability also increases.  The plots use a homogeneous damping value of $\alpha=1$.}
 \label{fig1g}
 \end{figure*}

The stability of the synchronous state of the uncontrolled system ($\gamma_j=0~\forall j$) is guaranteed if all eigenvalues $\mu_j$ of 
the Laplacian matrix are real and positive, see Eq. (\ref{eq:eigenvalue as function of laplacian}) and \cite{Manik2014,Motter2013}. If however
a given eigenvalue $\mu_j$ is negative, one of the corresponding eigenvalues $\lambda_{{j}{\pm}}$ is positive and the other one is negative; therefore, the synchronous state is unstable.
With added secondary control, i.e., $\gamma_j>0$, the region of stability increases, see  Fig. \ref{fig1g}, which holds for any number of nodes. Mathematically, the system is stable within the 
region defined by $\mu_{j} + \gamma_{j} > 0$, see also \cite{wang2016enhancing,Motter2013}.

Next, let us investigate the elementary system consisting of two nodes, a generator $(P_1>0)$ and a consumer $(P_2<0)$ first without secondary control to then investigate the benefits of adding such control.

\subsection{Uncontrolled two nodes system}
We consider a two nodes system without control, i.e., we set $\gamma_1=\gamma_2=0$, which is then governed by the following equations for the phase difference $\Delta \theta = \theta_1-\theta_2$ and the frequency difference
$\dot{\Delta \theta} = \omega{_1}-\omega_{2} = \Delta \omega$ with $\Delta P = P_1 - P_2$
\begin{equation}  
    \begin{array}{ll}
      \dot{\Delta \theta}=\Delta \omega,\\
      \dot{\Delta \omega}=-\alpha\Delta \omega + \Delta P -2K\sin(\Delta \theta),\\
     \end{array} \label{2nodest1} 
\end{equation}
where we have assumed homogeneous damping $\alpha_1 = \alpha_2 = \alpha$. 
The system has a steady state if and only if $2 K \geq \Delta P$, see also \cite{Manik2014}. The physical reason for the absence of a 
fixed point for $2 K< \Delta P$ is that the electric power flowing through a line cannot exceed the  maximal capacity $K$.\\
For $2 K > \Delta P$ the two steady states, $T_1$ and $T_2$, obtained from (\ref{2nodest1}), and their respective eigenvalues are \begin{equation} T_1: \left\{
  \begin{array}{ll}
      \Delta \theta^*=\arcsin\left(\frac{\Delta P}{2K} \right), \Delta \omega^{*}=0,\\
      \lambda_{\pm}(T_1) = -\frac{\alpha}{2} \pm \sqrt{\frac{\alpha}{2}^2 - \sqrt{4K^2 - \Delta P^2}}
      \end{array},
\right.
\label{T1}
\end{equation} \begin{equation}T_2: \left\{
  \begin{array}{ll}
      \Delta \theta^*=\pi-\arcsin\left(\frac{\Delta P}{2K} \right), \Delta \omega^{*}=0,\\
      \lambda_{\pm}(T_2) = -\frac{\alpha}{2} \pm \sqrt{\frac{\alpha}{2}^2 + \sqrt{4K^2 - \Delta P^2}}
  \end{array}.
\right.
\label{T2}
\end{equation} 
The steady state $T_1$ is a stable fixed point since we assume the damping $\alpha$ to be positive. In contrast, the steady state $T_2$ is a saddle since its eigenvalues $\lambda_{+}$ is a positive real number.\\
For $2K = \Delta P$, $T_1$ and $T_2$ collide via a saddle node bifurcation on a cycle (SNIC), entering a limit cycle  for $K < \frac{\Delta P}{2}$.
Such limit cycles often cause large frequency deviations that would result in the shutdown of (parts of) the grid and are therefore undesirable  \cite{Manik2014}.
But even for sufficient transmission capacity, i.e.  $2K \geq \Delta P$, the grid enters a limit cycle if we have unbalanced power, $P_1+P_2 \neq 0$ so that, from Eq.~\eqref{eq8} the synchronous angular velocity is given as 
\begin{equation}
 \omega_\text{syn}= \frac{P_1 + P_2}{2\alpha}.
 \label{synfr}
\end{equation}
Hence, the grid is no longer at its reference frequency of $f_R=50$ Hz or $60$ Hz but below it for $P_1+P_2<0$ and above it for $P_1+P_2>0$. To restore the frequency to the reference, we apply our secondary controller in the next subsection.

\subsection{Two nodes system with secondary control}
\begin{figure}[t]
\centering
     \includegraphics[scale=0.35] {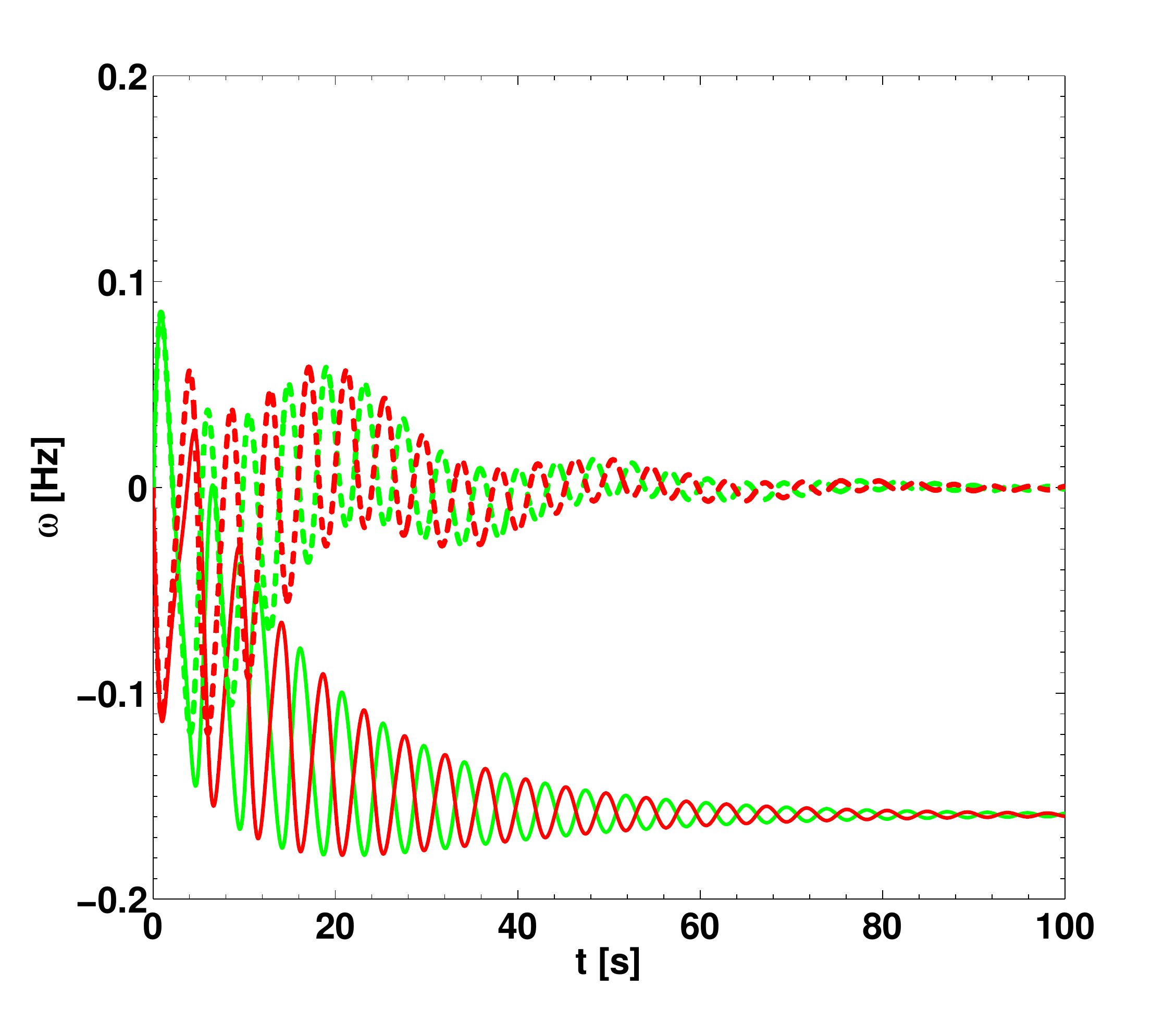} 
 \caption{\textbf{Including control restores the frequency back to the reference value.} We plot the time evolution of the angular velocity deviations $\omega$ without control (solid lines) and when controlling one node (dashed lines). With control, the system returns to $\omega=0$,
 i.e., the grid returns to its reference frequency $f_R$. Red and green curves
 represent the consumer and generator of a two node system respectively with parameters $\gamma=0.1~s^{-2}$,  
$\alpha=0.1~s^{-2}$, $K=1.5~s^{-2}$, $P_1=1~s^{-2}$, $P_2=-1.2~s^{-2}$.}
\label{fig21a}
\end{figure}

Next, we consider the two nodes system where one node applies a secondary control, i.e., we set the control parameters $\gamma_1=0$ and $\gamma_2=\gamma$ in the equation of motion \eqref{eq8}.
Then, the steady state of the controlled system 
is obtained as
\begin{equation*}
     \begin{array}{llll}
      \theta_1^*= \frac{P_{1} + P_{2}}{\gamma} + \arcsin(\frac{P_1}{K}), \\
      \theta_2^*=\frac{P_{1} + P_{2}}{\gamma},  \\
       \omega_1^*=0, \\
       \omega_2^*=0.
      \end{array}
\label{steady}
\end{equation*} 
For $P_1 > K$, there is no steady state and the system approaches a limit cycle, as the power cannot be transferred via the line and node 1 is uncontrolled. 
For $P_1 < K$ however, there will be a fixed point, even if the power is unbalanced $P_1+P_2 \neq 0$, in contrast to the uncontrolled system (Fig \ref{fig21a}).
While the uncontrolled system (solid lines) approaches a limit cycle with $\omega_{sync}$, as obtained by Eq. \eqref{synfr}, the controlled system is attracted to the fixed point, i.e. a stable operating state of the grid.

Next, we perform a stability analysis of the fixed point. 
Let $\mathbf{X}$=$(\delta\theta_1, \delta\theta_2, \delta\omega_1,\delta \omega_2)$ be a small perturbation of the fixed point.
The equations of motion of these small perturbations are given by
\begin{equation}
 \dot{\mathbf{X}}(t)=\mathbf{D} \cdot \mathbf{X}(t), 
 \label{compact}
\end{equation} where the matrix $D$ is defined as \begin{equation}
\mathbf{D}= \left[ {\begin{array}{cccc}
 0 & 0 & 1 & 0\\
 0 & 0 & 0 & 1\\
 -K\cos(\theta_1^* - \theta_2^*) & K\cos(\theta_1^* - \theta_2^*) & -\alpha & 0\\
  K\cos(\theta_1^* - \theta_2^*)  & -\gamma- K\cos(\theta_1^* - \theta_2^*) & 0& -\alpha
\end{array} } \right].
\end{equation} The characteristic polynomial of matrix $D$ is given as
\begin{equation}
 \lambda^4 + a_1\lambda^3 + a_2\lambda^2 + a_3\lambda + a_4=0,
 \label{polynomial}
\end{equation} where the parameters $a_1, a_2, a_3$ and $a_4$ are given by \begin{equation}
     \begin{array}{llll}
      a_1= 2\alpha,\\
      a_2= \alpha^2 + \gamma + 2a,\\
       a_3= 2a\alpha + \alpha\gamma,\\
       a_4= a\gamma,\\
       a=K\cos(\theta_1^* - \theta_2^*)= \sqrt{K^2 - P_{1}^2}.
      \end{array}
\label{par1}
\end{equation} 
To analyze the stability of the full four dimensional system, we need to obtain an expression for the eigenvalues. Unfortunately, a fourth or higher order polynomial does not have an easy to analyze solution so that we apply the Routh Hurwitz (RH)
criterion to determine the stability  \cite{xie2011criterion}. The RH criterion is a method which contains the necessary
and sufficient conditions for the stability of the system.
Given the polynomial  \begin{equation}
  P(\lambda)= \lambda^{n}+ a_{1}\lambda^{n-1}+...+a_{n-1}\lambda + a_{n},                                                                                            
\end{equation}
where the coefficients $a_{i}$ are real constants, $i=1,..,n$, we define the $n$ Hurwitz matrices
using the coefficients $a_{i}$ of the characteristic polynomial:
\begin{equation}
     \begin{array}{llll}
	B_1 &=&  (a_1),\\
	\vdots\\
	B_2 &=&  \left( {\begin{array}{cc}
 a_1 & 1\\
 a_3 & a_2 \end{array} } \right),\\
 B_n &= & \left( {\begin{array}{cccccc}
 a_{1} & 1 & 0 &   0   & \ddots & 0 \\
 a_{3} & a_{2} & a_{1} & 1 & \ddots & 0 \\
 a_{5} & a_{4} & a_{3} & a_{2} & \ddots & 0 \\
\vdots &\vdots &\vdots &\vdots & \ddots & \vdots \\
 0 & 0 & 0 & 0 & \ddots & a_{n} \end{array} } \right).
       \end{array}
\end{equation}
 According to the RH criterion, all 
 roots of the polynomial $P(\lambda)$ have negative real part 
 if and only if the determinant of all Hurwitz matrices are positive: $\det(B_{i}) > 0$, for all $i=1,2,...,n$ \cite{xie2011criterion}.
Applying the Routh Hurwitz criterion to the steady state of our two node system, we find that the steady state is stable if the following conditions are fulfilled: 
\begin{equation}
     \begin{array}{llll}
      a_1 > 0,\\
      a_3 > 0,\\
      a_4 > 0,\\
      a_1a_2a_3 - a_3^2 - a_1^2a_4 = d >0.
      \end{array}
\label{par2}
\end{equation} 
For the parameters used in this study, the three first conditions from \eqref{par2} are always fulfilled since $\alpha$, $\gamma$, $a>0$.
Hence, the steady state is stable if and only if $d>0$. In terms of the control parameter $\gamma$, we obtain the following inequality
\begin{equation}
 \alpha^2\gamma^2 + 2\alpha^{4}\gamma + 4a\alpha^{2}(a+\alpha^{2})>0,
 \label{cond4}
\end{equation} 
which again is always true; hence, as long as there is non-zero control, $\gamma>0$, the synchronous state is always stable, regardless of the further specific parameters of the system, highlighting the potential of secondary control.
Next, we shall investigate how secondary control interacts with changes of the network topology that lead to Braess' paradox in uncontrolled systems.
\begin{figure*}[ht!]
\centering
   \includegraphics[scale=0.5] {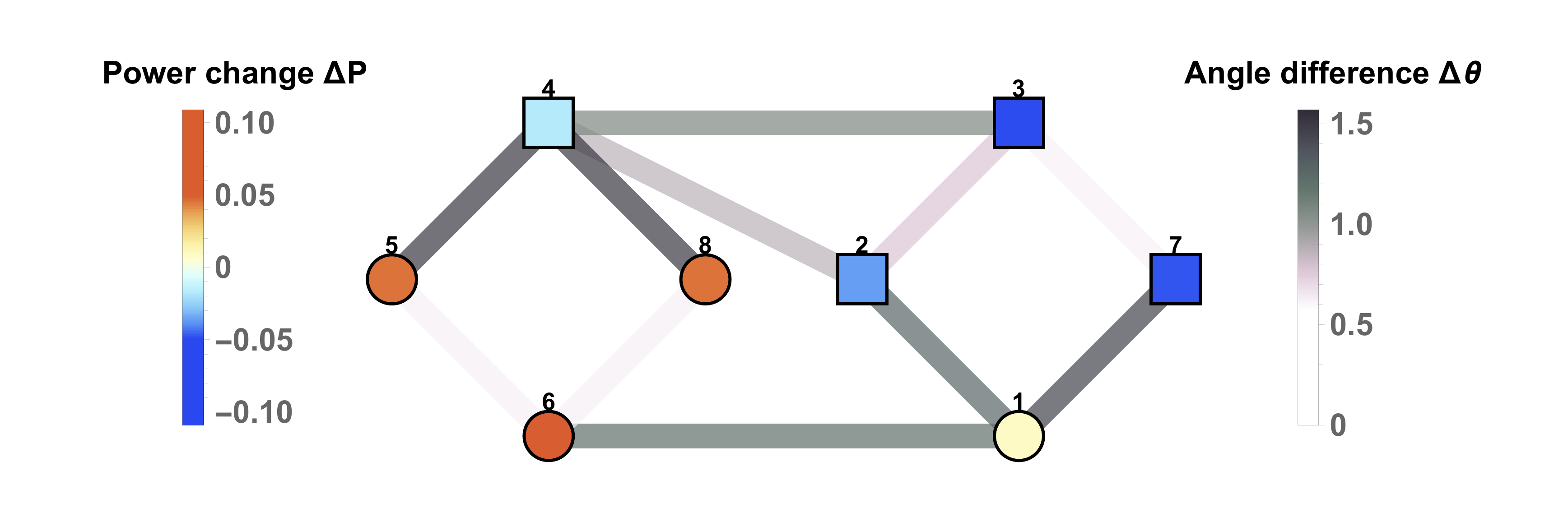} 
 \caption{Controlling all nodes changes the power dispatch in the system and allows stable operation without Braess' paradox.
We display the 8 node system susceptible to Braess' paradox with an added line $(2,4)$. 
Including secondary control causes all nodes to adapt their power following $\Delta P_i=-\gamma_i \theta_i$. This results in consumers (circles) to consume less (red: positive power change), while generators (squares) generate less (blue: negative power change). Thereby, the system preserves
its steady state even after including a line that causes an overload in the uncontrolled system. In addition, we note a very heterogeneous load of the lines (line color; darker colors indicate higher load). Specifically, the lines (4,5) and (4,8) are highly loaded, i.e., 
the phase difference $\Delta \theta_{4,5}$ becomes very large. Parameters used are
 $\alpha=1$ $s^{-1}$; $K=1.03$ $s^{-2}$, $\gamma=0.1$ $s^{-2}$.}
\label{fig4}
\end{figure*}

\begin{figure*}[ht!]
\centering
   \includegraphics[width=1.5\columnwidth] {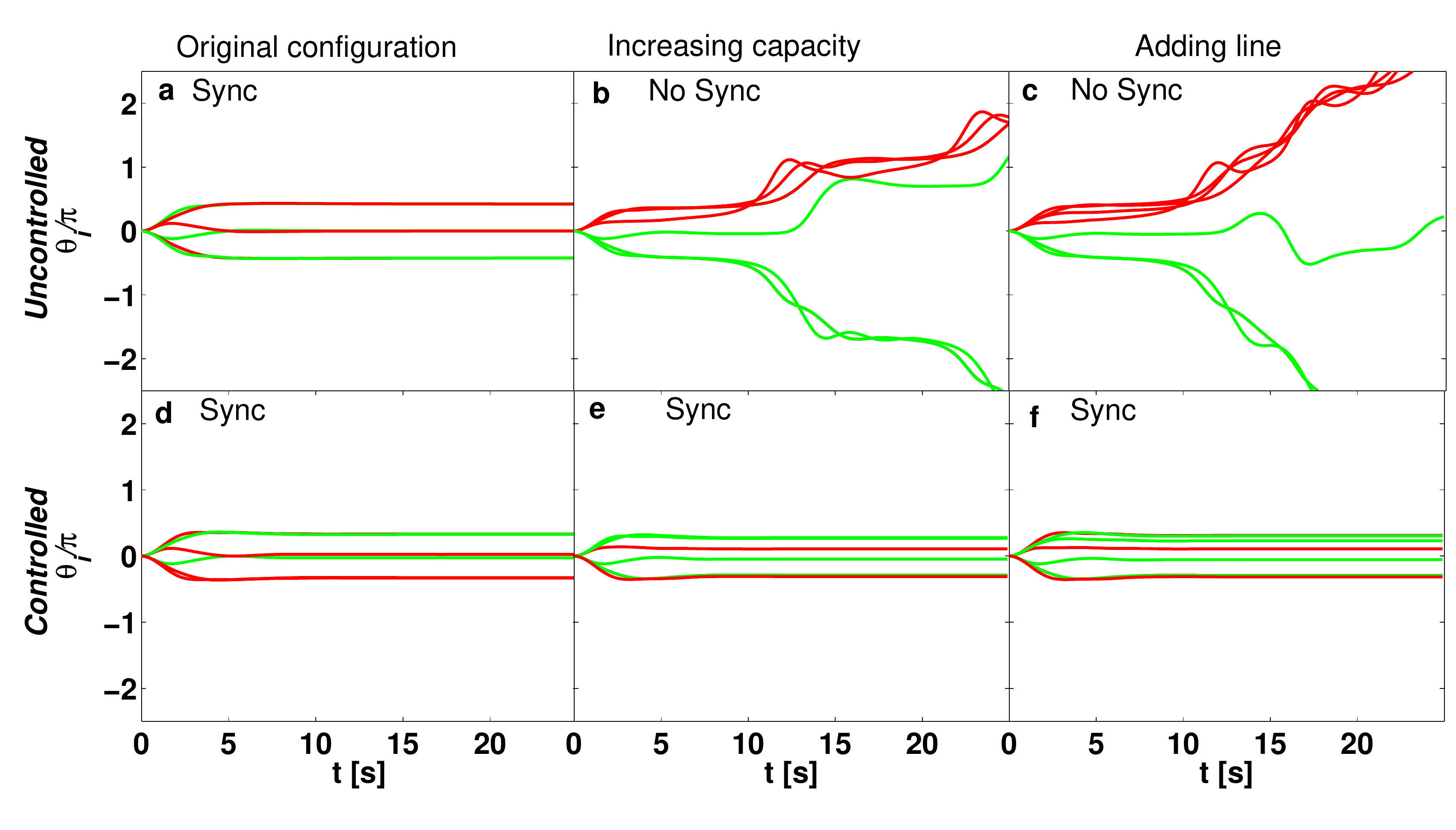} 
    \caption{Secondary control stabilizes a network after increasing capacity or adding a link. Braess' paradox in 
 power grids was observed when increasing the capacity of a line or adding an additional line caused the grid to lose its stable 
 fixed point (panels a-c), see also \cite{Witthaut2012}. In contrast,  applying secondary control guarantees stability (panels d-f).
We use the eight node system depicted in Fig. \ref{fig4}, only adding line $(2,4)$ for panels (c) and (f) and doubling the capacity of line $(3,4)$ in panels (b) and (e). Parameter are $\gamma=0.1~s^{-2}$, and $\alpha=1~s^{-1}$, $K=1.03~s^{-2}$ for all nodes and generator and consumer
power set to $P_{gen}=1~s^{-2}$, $P_{con}=-1~s^{-2}$, respectively. As in Fig.~\ref{fig21a}, red and green lines correspond to consumers and generators respectively.}
\label{fig3a}
\end{figure*}

\section{Braess' paradox prevented by secondary control}

Adding lines to a transmission network is intuitively expected to improve its synchronization ability. However, adding certain lines instead causes the grid to lose its synchronous state. More general, the effect of adding edges to a network and thereby causing problems and a decrease 
in performance was first predicted in 1968 for traffic networks \cite{Braess1968} and it is since known as Braess' paradox. It was observed in traffic systems in New York, USA \cite{Kolata1990}, and Stuttgart, Germany \cite{Knoedel2013}, when closing a street made the traffic go faster.\\
In electric networks, Braess' paradox has been predicted in DC power flow \cite{Labavic2014a}, AC power flow \cite{Motter2002} and recently in oscillator power grids \cite{Witthaut2012, Witthaut2013}. Building additional transmission capacity under specific conditions causes 
Braess' paradox and thereby the grid loses its fixed point and we observe a blackout.
Fortunately, not every network is susceptible to Braess' paradox. To study the effect of Braess' paradox in more detail, we investigate an elementary example  network composed of 8 nodes, where adding one additional transmission line or increasing the capacity of an existing line
leads to a
desynchronization of the network.  The network is shown in Fig. \ref{fig4} with an added line or in Fig. 1a in \cite{Witthaut2013}. \\
Braess' paradox is best understood when considering fixed point solutions. A fixed point exists for a power grid described by the swing equation,  if  $\dot{\theta}_i=\dot{\omega}_i=0$, which is equivalent to 
\begin{equation}
\sum_{j=1}^{N} K_{ij}\sin(\theta_i - \theta_j)=P_i~\forall i\in {1,...,N},
\label{eq:fixedPointCondition}
\end{equation}
if we set $\gamma_i=0$. These algebraic equations do not always have a solutions for the phases $\theta_i$. As a trivial example, without enough transmission capacity, i.e, $\sum_{j=1}^{N} K_{ij} < P_{i}$ for finite power $P_i\neq0$ there cannot be any fixed point.\\
In addition, adding a line in a network can result in the equations to be overdetermined and therefore to have no solution.
Without solution, there is no fixed point and the grid desynchronizes in the absence of adequate counter measures (e.g. by external controllers), see Fig. \ref{fig3a}a, b and c and \cite{Witthaut2012}. We may also interpret these results in the light of the cirtical coupling $K_c$ of 
the grid \cite{Manik2014}. The critical coupling is defined as the minimum value of $K$ so that for a homogeneously coupled grid, i.e. $K_{ij}=K k_{ij}$ with unweighted adjacency matrix $\mathbf{k}$, the algebraic equations \eqref{eq:fixedPointCondition} have at least one solution. Thereby, $K_c$ gives the minimum capacity necessary to synchronize the grid.
Adding a line or increasing the capacity of one existing line effectively increases the critical coupling $K_c$  grid \cite{Witthaut2012}.
 Increasing $K_c$ means the fixed point can only be restored by increasing the capacity $K$ for all lines.\\
Is Braess' paradox still present after adding secondary control? Let us consider again the 8 node network (see Fig. \ref{fig4} for the network with added line). However, we add our PD controller in each node, such that the equations of motion are governed by
\begin{equation}  
    \begin{array}{ll}
      \dot{\mathbf{\theta}}_i={\omega}_{i},\\
      \dot{\omega}_i=-\alpha_{i}\dot{\mathbf{\theta}}_{i} -\gamma_{i}{\mathbf{\theta}}_{i} + {P_{i}} - \sum_{j=1}^{8}K_{ij}\sin(\theta_i - \theta_j).\\
     \end{array}
 \label{8cnode} 
\end{equation}

Adding the control on all nodes reliably prevents Braess' paradox, see Fig. \ref{fig3a}. There, we plot the phase of each node, with consumers in red and generators in green, using both the uncontrolled (first row) 
and the controlled (second row) network in three configurations: original network, increasing capacity and adding a new link. Specifically, we double the capacity of the edge $(3,4)$ or add the line $(2,4)$.
The original network is stable regardless whether it is controlled or not. It enters a phase-locked state where all machines run in synchrony (Fig. \ref{fig3a}a and d).
However, the steady state is lost when the capacity of a line is increased or  a new link is added to the network (Fig. \ref{fig3a}b and c). 
On the other hand, controlling the network guarantees a stable state even after increasing the capacity or adding a line, thereby preventing Braess' paradox (Fig. \ref{fig3a}e and f). 
\\
To better understand how the controller stabilizes the network, we note that the effective power generated/consumed at each node is given by
\begin{equation}
\label{eq:PowerChange}
P_{i}^\text{eff}=P_i -\gamma_i \theta_i.
\end{equation}
In case of the 8 node system, we plot the change of power $\Delta P_i=-\gamma \theta_i$ in Fig. \ref{fig4}.  With control, generators have negative and consumers positive power change, i.e., the total consumption and the total generation are decreased. The condition to find a fixed point,
Eq. \eqref{eq:fixedPointCondition}, changes with added control to 
\begin{equation}
\sum_{j=1}^{N} K_{ij}\sin(\theta_i - \theta_j)=P_i-\gamma_i \theta_i~\forall i\in {1,...,N}.
\label{eq:fixedPointConditionControl}
\end{equation}
For all configurations investigated, this equation has a solution if $\gamma_i > 0~\forall i$. Even if the transmission capacity $K_{ij}$ is insufficient or would normally cause Braess' paradox, the term $-\gamma_i {\theta}$ balances the equation and guarantees a solution. 
Thereby, we do not need to increase the capacity of all lines because the control reduces the total load in the system.

We illustrate this for a 2 node system with $\gamma_1=\gamma_2=\gamma$. The critical coupling is then given as 
\begin{equation}
K_{c}^{New}=K_{c}-\gamma\theta,
\label{ste4} 
\end{equation}
i.e. the controller reduces the load on the lines, enabling the fixed point with lower capacity. Following the same argumentation, secondary control also cures Braess' paradox which would otherwise require an increase of the transmission capacity.

\begin{figure*}
\centering
\includegraphics[width=0.9\textwidth] {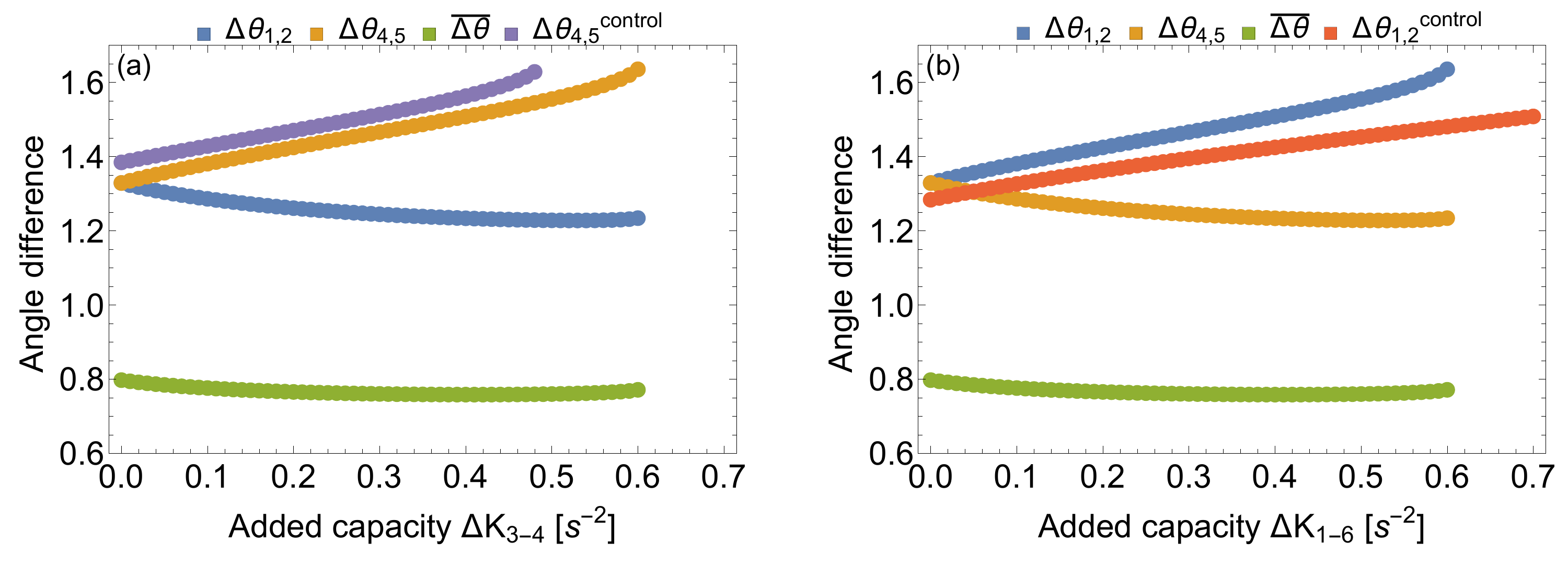} 
 \caption{
Effectiveness of secondary control in curing Braess' paradox depends on the topology.
 Here we plot the phase differences between the nodes connected by the most loaded lines and the average phase difference when increasing the capacity of lines  $(3,4)$ in panel a and  $(1,6)$ in panel b, for the grid shown in Fig.  \ref{fig4} but without the line $(2, 4)$.
We plot the angle differences up to the critical added coupling $\Delta K_c$ for which there is no longer any fixed point. The uncontrolled system fails at $\Delta K_c\approx 0.6$.
 Adding control only on the generator nodes increases the load on line  $(4, 5) $. Thereby, we observe an earlier failure ($\Delta K_c\approx 0.5$) with control when modifying line  $(3,4 )$ in panel a, compared to the uncontrolled case. Contrary, the control reduces  the load on line  $(1,2) $ and the grid does not display Braess' paradox any more, when modifying line  $(1,6 )$ as shown in panel b. 
  Parameters are $\alpha=1 s^{-1}$; $K=1.03 s^{-2}$ for all lines except the one with added capacity and $\gamma^\text{Generator}=0.1 s^{-2}$.
}
\label{AngleDifferenceVsDeltaK}
\end{figure*}

So far, we assumed that we control all nodes in the network. Consumer nodes, however, may have limited generation capacity and therefore limited control capability. Therefore, let us now assume that control is only available at the nodes with positive power generation (generators), as usual in today's power grids \cite{Machowski2011}. In this case, the effectiveness of the control depends strongly on the topology, e.g. which line is getting upgraded. We consider two cases.

\begin{figure*}
\centering
   \includegraphics[width=0.9\textwidth] {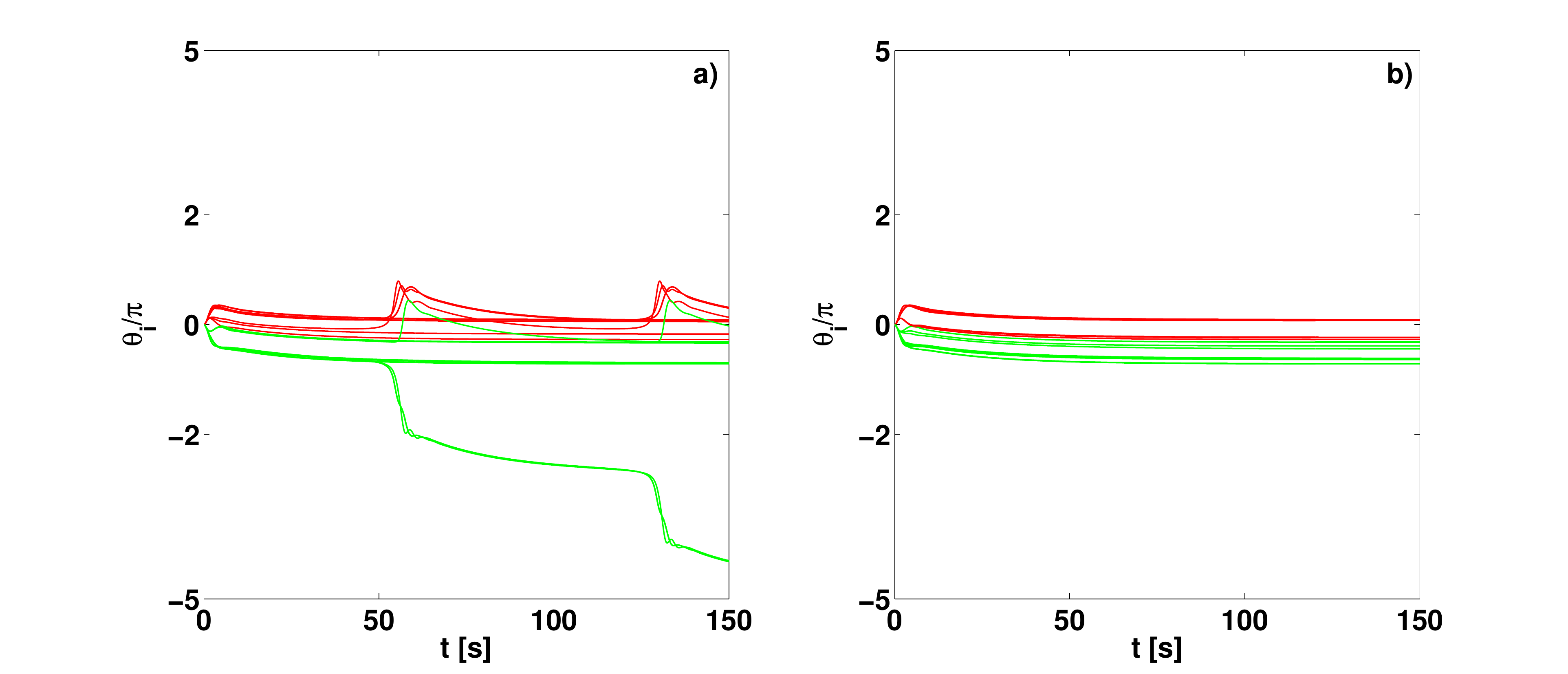} 
 \caption{Controlling generators does not reliably prevent Braess' paradox. We show the time evolution of the phases of the nodes for the grid as shown in Fig. \ref{fig4} but without line $(2,4)$ after increasing the capacity of one line, specifically (a):  $\Delta K_{3,4}=1.0 s^{-2}$ or (b) $\Delta K_{1,6}=1.0 s^{-2}$. 
When adding capacity to line $(3,4)$, the control on the generators cannot prevent a loss of the fixed point (panel a). However, the angles do not diverge as drastically as in an uncontrolled case (compare Fig. \ref{fig3a} c). Contrary, applying control only on the generators fully prevents Braess' paradox, when line $(1,6)$ is modified.
 Parameters are $\alpha=1 s^{-1}$; $K=1.03 s^{-2}$, $\gamma^\text{Generator}=0.1 s^{-2}$ and green lines show the dynamics of consumers with red lines giving the generator dynamics.
 }
\label{controlgen}
\end{figure*}

First, we increase the capacity of line $(3,4)$ between two generator nodes by $\Delta K$, which without control eventually leads to Braess' Paradox (Fig. \ref{fig3a}b). As a measure of the stability of the system, we evaluate the phase differences $\Delta\theta_{i,j}=\theta_i-\theta_j$, as a measure of the load of the line connecting nodes $i$ and $j$. These phase differences are obtained from the fixed point given by Eq. (\ref{eq:fixedPointConditionControl}), computed by Newton' method. Without control, the phase difference $\Delta \theta_{4,5}$, which is the same as $\Delta \theta_{4,8}$, increases continuously with $\Delta K$ while the average phase difference $\overline{\Delta \theta}$ stays almost constant (see Fig. \ref{AngleDifferenceVsDeltaK}a). Eventually, at $\Delta K \sim 0.6 s^{-2}$, power lines can not deliver the necessary power to some nodes and there is no longer a fixed point. The system enters an oscillatory regime, as shown in Fig. \ref{fig3a}. Adding control does not help to improve the situation. In fact the phase difference $\Delta \theta_{4,5}$  increases faster with $\Delta K$ and the fixed point disappears for even lower values of $\Delta K$, at $\Delta K \sim 0.49 s^{-2}$. So, controlling only the generators does not prevent Braess' paradox in this case. Nevertheless, the oscillatory regime reached after the instability is somehow different in the cases with and without control. As shown in Fig. \ref{controlgen}a.  With control most of the nodes remain synchronized at the reference frequency and only two nodes show phase slips at a slow time scale. On the contrary, without control all nodes rotate showing phase slips (see Fig. \ref{fig3a}b for the same $\Delta K$).

Next, we increase the capacity of line $(1,6)$, connecting two consumer nodes, by $\Delta K$. Without control, increasing $\Delta K$  the phase difference $\Delta \theta_{1,2}$ increases continuously (see Fig. \ref{AngleDifferenceVsDeltaK}b) until the system become unstable at $\Delta K \sim 0.6 s^{-2}$ leading to Braess' paradox. Applying control exclusively to generator nodes does indeed help in this case. The range of existence of the fixed point is extended to any value of $ \Delta K$ preventing the paradox completely, see Fig. \ref{controlgen}.

\section{Discussion}

Concluding, we have studied a simple secondary control algorithm that successfully restores the grid frequency of an unbalanced power and  may prevent Braess' paradox. 

Secondary control, when applied to all nodes, improves the stability of the grid, regardless of topology, and even allows stable operation for mismatched power \cite{Machowski2011,Wood2013,Kundur1994}.
While primary control stabilizes the frequency, secondary control restores the frequency to the reference value and always guarantees the existence of a stable fixed point.
We have systematically computed the fixed point stability of the power grid with secondary control as a function of both the network topology and the control action.
Thereby, we have extended previous stability analysis of uncontrolled systems \cite{Motter2013} or systems including secondary control restricted to balanced power \cite{Dongmo2017}.

Secondary control in all nodes may also  prevent the loss of the operational state via Braess' paradox. As shown by Witthaut and Timme \cite{Witthaut2012,Witthaut2013}, the addition of certain transmission lines may lead to a loss of the operational state of the power grid. Using primary control 
only \cite{Manik2014}, does not suffice to prevent Braess' paradox.
In contrast, we have now demonstrated that secondary control prevents the desynchronization in networks prone to Braess' paradox if all nodes, i.e., consumers and generators alike, are controlled (Figs. \ref{fig4} and \ref{fig3a}). 
The control reduces the total amount of net power generated and consumed at each node of the grid guaranteeing that the transmitted power does not exceed the transmission capacity.
Thereby, it offers a trade-off between grid extension and investments in control, assuming some amount of local generation is possible.

In today's grid, secondary control is implemented only in power plants. Thus nodes with generation much larger than consumption, i.e., generator nodes, have a large control capability while nodes in which consumption is larger than generation, consumer nodes, have very little, if any, control capability. If control does only takes place at generator nodes, its efficiency strongly depends on the grid topology. We have observed that if the capacity of a line connecting two generator nodes is increased, the control does not prevent  Braess' paradox. On the contrary, in the case of increasing the capacity of a line connecting two consumer nodes, secondary control is capable of redistributing the power flow so that lines are not overloaded and the paradox is avoided.

Concluding, using secondary control on all nodes in a network improves its stability and robustness with respect to dynamical and topological perturbations. If control is mainly available in generator nodes only the effectiveness of the control depends strongly on the topology of the network. This stresses the importance of involving consumers, e.g. via demand control schemes or local generation (prosumers) in future grids \cite{Fang2012,Schaefer2015,Schaefer2016}. Finally, further research is necessary to extend our results, e.g., to alternative control mechanisms. One example is to allow $\tau>0$ in Eq. \eqref{eq7}, i.e., making the power provided by each node explicitly time-dependent.

\begin{acknowledgments}
We gratefully acknowledge support from the Federal Ministry of Education
and Research (BMBF grant no.03SF0472A-F to B.S., M.T. and D.W.),
the Göttingen Graduate School for Neurosciences and Molecular Biosciences
(DFG Grant GSC 226/2 to B.S.), the Max Planck Society (to M.T.), the Helmholtz
Association (via the joint initiative ``Energy System 2050 - A Contribution
of the Research Field Energy'' and the grant no.VH-NG-1025 to D.W.),
the Agencia Estatal de Investigaci\'on (AEI, Spain) 
and Fondo Europeo de Desarrollo Regional under Project ESoTECoS, grant 
numbers FIS2015-63628-C2-1-R (AEI/FEDER,UE) and FIS2015-63628-C2-2-R 
(AEI/FEDER,UE). E.B.T.-T. also acknowledges the fellowship 
FIS2015-63628-CZ-Z-R under the FPI program of MINEICO, Spain.
\end{acknowledgments}

%

\end{document}